\documentstyle[seceq,twocolumn]{jpsj}       
\edef\psfigRestoreAt{\catcode`@=\number\catcode`@\relax}
\catcode`\@=11\relax
\newwrite\@unused
\def\typeout#1{{\let\protect\string\immediate\write\@unused{#1}}}
\def\figurepath{./}

\def\@nnil{\@nil}
\def\@empty{}
\def\@psdonoop#1\@@#2#3{}
\def\@psdo#1:=#2\do#3{\edef\@psdotmp{#2}\ifx\@psdotmp\@empty \else
    \expandafter\@psdoloop#2,\@nil,\@nil\@@#1{#3}\fi}
\def\@psdoloop#1,#2,#3\@@#4#5{\def#4{#1}\ifx #4\@nnil \else
       #5\def#4{#2}\ifx #4\@nnil \else#5\@ipsdoloop #3\@@#4{#5}\fi\fi}
\def\@ipsdoloop#1,#2\@@#3#4{\def#3{#1}\ifx #3\@nnil 
       \let\@nextwhile=\@psdonoop \else
      #4\relax\let\@nextwhile=\@ipsdoloop\fi\@nextwhile#2\@@#3{#4}}
\def\@tpsdo#1:=#2\do#3{\xdef\@psdotmp{#2}\ifx\@psdotmp\@empty \else
    \@tpsdoloop#2\@nil\@nil\@@#1{#3}\fi}
\def\@tpsdoloop#1#2\@@#3#4{\def#3{#1}\ifx #3\@nnil 
       \let\@nextwhile=\@psdonoop \else
      #4\relax\let\@nextwhile=\@tpsdoloop\fi\@nextwhile#2\@@#3{#4}}
\newread\ps@stream
\newif\ifnot@eof       % continue looking for the bounding box?
\newif\if@noisy        % report what you're making?
\newif\if@atend        % %%BoundingBox: has (at end) specification
\newif\if@psfile       % does this look like a PostScript file?
{\catcode`\%=12\global\gdef\epsf@start{%!}}
\def\epsf@PS{PS}
\def\epsf@getbb#1{%
\openin\ps@stream=#1
\ifeof\ps@stream\typeout{Error, File #1 not found}\else
   {\not@eoftrue \chardef\other=12
    \def\do##1{\catcode`##1=\other}\dospecials \catcode`\ =10
    \loop
       \if@psfile
	  \read\ps@stream to \epsf@fileline
       \else{
	  \obeyspaces
          \read\ps@stream to \epsf@tmp\global\let\epsf@fileline\epsf@tmp}
       \fi
       \ifeof\ps@stream\not@eoffalse\else
       \if@psfile\else
       \expandafter\epsf@test\epsf@fileline:. \\%
       \fi
          \expandafter\epsf@aux\epsf@fileline:. \\%
       \fi
   \ifnot@eof\repeat
   }\closein\ps@stream\fi}%
\long\def\epsf@test#1#2#3:#4\\{\def\epsf@testit{#1#2}
			\ifx\epsf@testit\epsf@start\else
\typeout{Warning! File does not start with `\epsf@start'.  It may not be a PostScript file.}
			\fi
			\@psfiletrue} % don't test after 1st line
{\catcode`\%=12\global\let\epsf@percent=%\global\def\epsf@bblit{%BoundingBox}}
\long\def\epsf@aux#1#2:#3\\{\ifx#1\epsf@percent
   \def\epsf@testit{#2}\ifx\epsf@testit\epsf@bblit
	\@atendfalse
        \epsf@atend #3 . \\%
	\if@atend	
	   \if@verbose{
		\typeout{psfig: found `(atend)'; continuing search}
	   }\fi
        \else
        \epsf@grab #3 . . . \\%
        \not@eoffalse
        \global\no@bbfalse
        \fi
   \fi\fi}%
\def\epsf@grab #1 #2 #3 #4 #5\\{%
   \global\def\epsf@llx{#1}\ifx\epsf@llx\empty
      \epsf@grab #2 #3 #4 #5 .\\\else
   \global\def\epsf@lly{#2}%
   \global\def\epsf@urx{#3}\global\def\epsf@ury{#4}\fi}%
\def\epsf@atendlit{(atend)} 
\def\epsf@atend #1 #2 #3\\{%
   \def\epsf@tmp{#1}\ifx\epsf@tmp\empty
      \epsf@atend #2 #3 .\\\else
   \ifx\epsf@tmp\epsf@atendlit\@atendtrue\fi\fi}

\def\psdraft{
	\def\@psdraft{0}
}
\def\psfull{
	\def\@psdraft{100}
}

\psfull

\newif\if@draftbox
\def\psnodraftbox{
	\@draftboxfalse
}
\@draftboxtrue

\newif\if@prologfile
\newif\if@postlogfile
\def\pssilent{
	\@noisyfalse
}
\def\psnoisy{
	\@noisytrue
}
\pssilent % 970825 HGE
\newif\if@bbllx
\newif\if@bblly
\newif\if@bburx
\newif\if@bbury
\newif\if@height
\newif\if@width
\newif\if@rheight
\newif\if@rwidth
\newif\if@clip
\newif\if@verbose
\def\@p@@sclip#1{\@cliptrue}

\def\@p@@sfile#1{\def\@p@sfile{null}%
	        \openin1=#1
		\ifeof1\closein1%
		       \openin1=\figurepath#1
			\ifeof1\typeout{Error, File #1 not found}
			\else\closein1
			    \edef\@p@sfile{\figurepath#1}%
                        \fi%
		 \else\closein1%
		       \def\@p@sfile{#1}%
		 \fi}
\def\@p@@sfigure#1{\def\@p@sfile{null}%
	        \openin1=#1
		\ifeof1\closein1%
		       \openin1=\figurepath#1
			\ifeof1\typeout{Error, File #1 not found}
			\else\closein1
			    \def\@p@sfile{\figurepath#1}%
                        \fi%
		 \else\closein1%
		       \def\@p@sfile{#1}%
		 \fi}

\def\@p@@sbbllx#1{
		\@bbllxtrue
		\dimen100=#1
		\edef\@p@sbbllx{\number\dimen100}
}
\def\@p@@sbblly#1{
		\@bbllytrue
		\dimen100=#1
		\edef\@p@sbblly{\number\dimen100}
}
\def\@p@@sbburx#1{
		\@bburxtrue
		\dimen100=#1
		\edef\@p@sbburx{\number\dimen100}
}
\def\@p@@sbbury#1{
		\@bburytrue
		\dimen100=#1
		\edef\@p@sbbury{\number\dimen100}
}
\def\@p@@sheight#1{
		\@heighttrue
		\dimen100=#1
   		\edef\@p@sheight{\number\dimen100}
}
\def\@p@@swidth#1{
		\@widthtrue
		\dimen100=#1
		\edef\@p@swidth{\number\dimen100}
}
\def\@p@@srheight#1{
		\@rheighttrue
		\dimen100=#1
		\edef\@p@srheight{\number\dimen100}
}
\def\@p@@srwidth#1{
		\@rwidthtrue
		\dimen100=#1
		\edef\@p@srwidth{\number\dimen100}
}
\def\@p@@ssilent#1{ 
		\@verbosefalse
}
\def\@p@@sprolog#1{\@prologfiletrue\def\@prologfileval{#1}}
\def\@p@@spostlog#1{\@postlogfiletrue\def\@postlogfileval{#1}}
\def\@cs@name#1{\csname #1\endcsname}
\def\@setparms#1=#2,{\@cs@name{@p@@s#1}{#2}}
\def\ps@init@parms{
		\@bbllxfalse \@bbllyfalse
		\@bburxfalse \@bburyfalse
		\@heightfalse \@widthfalse
		\@rheightfalse \@rwidthfalse
		\def\@p@sbbllx{}\def\@p@sbblly{}
		\def\@p@sbburx{}\def\@p@sbbury{}
		\def\@p@sheight{}\def\@p@swidth{}
		\def\@p@srheight{}\def\@p@srwidth{}
		\def\@p@sfile{}
		\def\@p@scost{10}
		\def\@sc{}
		\@prologfilefalse
		\@postlogfilefalse
		\@clipfalse
		\if@noisy
			\@verbosetrue
		\else
			\@verbosefalse
		\fi
}
\def\parse@ps@parms#1{
	 	\@psdo\@psfiga:=#1\do
		   {\expandafter\@setparms\@psfiga,}}
\newif\ifno@bb
\def\bb@missing{
	\if@verbose{
		\typeout{psfig: searching \@p@sfile \space  for bounding box}
	}\fi
	\no@bbtrue
	\epsf@getbb{\@p@sfile}
        \ifno@bb \else \bb@cull\epsf@llx\epsf@lly\epsf@urx\epsf@ury\fi
}	
\def\bb@cull#1#2#3#4{
	\dimen100=#1 bp\edef\@p@sbbllx{\number\dimen100}
	\dimen100=#2 bp\edef\@p@sbblly{\number\dimen100}
	\dimen100=#3 bp\edef\@p@sbburx{\number\dimen100}
	\dimen100=#4 bp\edef\@p@sbbury{\number\dimen100}
	\no@bbfalse
}
\def\compute@bb{
		\no@bbfalse
		\if@bbllx \else \no@bbtrue \fi
		\if@bblly \else \no@bbtrue \fi
		\if@bburx \else \no@bbtrue \fi
		\if@bbury \else \no@bbtrue \fi
		\ifno@bb \bb@missing \fi
		\ifno@bb \typeout{FATAL ERROR: no bb supplied or found}
			\no-bb-error
		\fi
		\count203=\@p@sbburx
		\count204=\@p@sbbury
		\advance\count203 by -\@p@sbbllx
		\advance\count204 by -\@p@sbblly
		\edef\@bbw{\number\count203}
		\edef\@bbh{\number\count204}
}
\def\in@hundreds#1#2#3{\count240=#2 \count241=#3
		     \count100=\count240	% 100 is first digit #2/#3
		     \divide\count100 by \count241
		     \count101=\count100
		     \multiply\count101 by \count241
		     \advance\count240 by -\count101
		     \multiply\count240 by 10
		     \count101=\count240	%101 is second digit of #2/#3
		     \divide\count101 by \count241
		     \count102=\count101
		     \multiply\count102 by \count241
		     \advance\count240 by -\count102
		     \multiply\count240 by 10
		     \count102=\count240	% 102 is the third digit
		     \divide\count102 by \count241
		     \count200=#1\count205=0
		     \count201=\count200
			\multiply\count201 by \count100
		 	\advance\count205 by \count201
		     \count201=\count200
			\divide\count201 by 10
			\multiply\count201 by \count101
			\advance\count205 by \count201
		     \count201=\count200
			\divide\count201 by 100
			\multiply\count201 by \count102
			\advance\count205 by \count201
		     \edef\@result{\number\count205}
}
\def\compute@wfromh{
		\in@hundreds{\@p@sheight}{\@bbw}{\@bbh}
		\edef\@p@swidth{\@result}
}
\def\compute@hfromw{
		\in@hundreds{\@p@swidth}{\@bbh}{\@bbw}
		\edef\@p@sheight{\@result}
}
\def\compute@handw{
		\if@height 
			\if@width
			\else
				\compute@wfromh
			\fi
		\else 
			\if@width
				\compute@hfromw
			\else
				\edef\@p@sheight{\@bbh}
				\edef\@p@swidth{\@bbw}
			\fi
		\fi
}
\def\compute@resv{
		\if@rheight \else \edef\@p@srheight{\@p@sheight} \fi
		\if@rwidth \else \edef\@p@srwidth{\@p@swidth} \fi
}
\def\compute@sizes{
	\compute@bb
	\compute@handw
	\compute@resv
}
\def\psfig#1{\vbox {
	\ps@init@parms
	\parse@ps@parms{#1}
	\compute@sizes
	\ifnum\@p@scost<\@psdraft{
		\if@verbose{
			\typeout{psfig: including \@p@sfile \space }
		}\fi
		\special{ps::[begin] 	\@p@swidth \space \@p@sheight \space
				\@p@sbbllx \space \@p@sbblly \space
				\@p@sbburx \space \@p@sbbury \space
				startTexFig \space }
		\if@clip{
			\if@verbose{
				\typeout{(clip)}
			}\fi
			\special{ps:: doclip \space }
		}\fi
		\if@prologfile
		    \special{ps: plotfile \@prologfileval \space } \fi
		\special{ps: plotfile \@p@sfile \space }
		\if@postlogfile
		    \special{ps: plotfile \@postlogfileval \space } \fi
		\special{ps::[end] endTexFig \space }
		\vbox to \@p@srheight true sp{
			\hbox to \@p@srwidth true sp{
				\hss
			}
		\vss
		}
	}\else{
		\if@draftbox{		
			\hbox{\fbox{\vbox to \@p@srheight true sp{
			\vss
			\hbox to \@p@srwidth true sp{ \hss \@p@sfile \hss }
			\vss
			}}}
		}\else{
			\vbox to \@p@srheight true sp{
			\vss
			\hbox to \@p@srwidth true sp{\hss}
			\vss
			}
		}\fi

	}\fi
}}
\def\psglobal{\typeout{psfig: PSGLOBAL is OBSOLETE; use psprint -m instead}}
\psfigRestoreAt

\newcommand{\MF}{{\large{\manual META}\-{\manual FONT}}}
\newcommand{\manual}{rm}        % Substitute rm (Roman) font.
\newcommand\bs{\char '134 }     % add backslash char to \tt font

\newcommand{\scr}[1]{\mbox{\scriptsize #1}}
\newcommand{\ov}[1]{\overline{#1}}
\newcommand{\capt}[2]{\caption[#1]{\footnotesize #2}}
\newcommand{\dt}{\Delta \tau}
\newcommand{\eff}{_{\mbox{\scriptsize eff}}}
\newcommand{\PtIr}{${\rm Sr}_3 {\rm CuPt}_{1-x} {\rm Ir}_x {\rm O}_6$}
\newcommand{\Cu}{${\rm Cu}$}
\newcommand{\Ir}{${\rm Ir}$}
\newcommand{\Pt}{${\rm Pt}$}

\def\runtitle{Monte Carlo Study of the Separation of Energy Scales}
\def\runauthor{Beat {\sc Ammon} and Manfred {\sc Sigrist}}

\title{
Monte Carlo Study of the Separation of Energy Scales in 
Quantum Spin 1/2 Chains with Bond Disorder
}

\author{
Beat {\sc Ammon}\footnote{Present address:  Institute for Solid State Physics, 
	University of Tokyo, Roppongi 7-22-1, Tokyo 106. 
     E-mail: ammon@ginnan.issp.u-tokyo.ac.jp}
and
Manfred {\sc Sigrist}$^{1,}$\footnote{E-mail: sigrist@yukawa.kyoto-u.ac.jp}
}
\inst{
SCSC and Theoretische Physik, 
Eidgen\"ossische Technische Hochschule, CH-8092 Z\"urich, Switzerland\\
$^1$Yukawa Institute for Theoretical Physics, Kyoto University,
Kyoto 606-8502	
}

\recdate{
}

\abst{
  One-dimensional Heisenberg spin $1/2$ chains with random ferro- and
  antiferromagnetic bonds are realized in systems such as \PtIr. We
  have investigated numerically the thermodynamic properties of a
  generic random bond model and of a realistic model of \PtIr \mbox{}
  by the quantum Monte Carlo loop algorithm. For the first time we
  demonstrate the separation into three different temperature regimes
  for the original Hamiltonian based on an exact treatment, especially
  we show that the intermediate temperature regime is well-defined and
  observable in both the specific heat and the magnetic
  susceptibility. The crossover between the regimes is indicated by
  peaks in the specific heat. The uniform magnetic susceptibility
  shows Curie-like behavior in the high-, intermediate- and
  low-temperature regime, with different values of the Curie constant
  in each regime. We show that these regimes are overlapping in the
  realistic model and give numerical data for the analysis of
  experimental tests.  }

\kword
{
quasi-one-dimensional spin system, random spin chain, quantum Monte Carlo,
${\rm Sr}_3 {\rm CuPt}_{1-x} {\rm Ir}_x {\rm O}_6$
}

\begin{document}
\sloppy
\maketitle
%%%%%%%%%%%%%%%%%%%%%%%%%%%%%%%%%%%%%%%%%%%%%%%%%%%%%%%%%%%%%%%%%%%%%%%%
% INTRODUCTION
%%%%%%%%%%%%%%%%%%%%%%%%%%%%%%%%%%%%%%%%%%%%%%%%%%%%%%%%%%%%%%%%%%%%%%%%
%****************************************************************
\section{Introduction}
%****************************************************************

One-dimensional quantum spin chains are typical examples of many-body
systems with a very rich variety of physical properties. Over many
decades they have attracted much interest in theory and have motivated
the development of various calculation schemes, both analytical and
numerical. The number of real compounds containing
quasi-one-dimensional spin systems is growing. A few examples are
organic systems like NENP and NINO \cite{nenp,nino} or inorganic
compounds such as ${\rm Sr}_3 {\rm MPtO}_6 $ (M = Ni, Cu and Zn)
\cite{wilkinson}. Some of them are investigated as possible
realizations of Haldane gap systems. Recently also ladder systems such
as $ ({\rm Ca})_2 {\rm V}{\rm O}_3 $ and ${\rm Sr}_2 {\rm CuO}_3 $
were investigated as an example of a resonating valence bond system.
All these systems are supposed to be regular chains or ladders.
However, in practice disorder must occur in most of these systems.
Already little disorder in the composition can have considerable
influence on the low-energy properties. A peculiar example of a
disordered spin chain was recently discovered by Nguyen and zur Loye:
the alloy $ {\rm Sr}_3 {\rm CuPt}_{1-x} {\rm Ir}_x {\rm O}_6
$\cite{zur_loye}. The pure compound $ {\rm Sr}_3 {\rm CuPtO}_6 $ forms
an antiferromagnetic (AF) spin chain where a spin 1/2 is provided by
each Cu-ion. In this compound the Cu-ions are alternating with the
spinless Pt-ions along chains. If all Pt is replaced by Ir which
carries a spin 1/2, then the system is a ferromagnetic (FM) spin
chain. Therefore the random alloy $ {\rm Sr}_3 {\rm CuPt}_{1-x} {\rm
  Ir}_x {\rm O}_6 $ is a spin system with two types of bonds, FM ($
J_{\mbox{\scriptsize F}} <0 $) and AF ($ J_{\mbox{\scriptsize A}} >0 $), 
which are randomly distributed. There
is a correlation among the bonds in the sense that FM bonds always
occur in sequences of even numbers, since each Ir-ion makes such bonds
with its two neighboring Cu-ions. A generic model to study the
properties of such a system was considered by Furusaki and coworkers
\cite{Furusaki} who neglected the correlation among the bonds and used
$J=|J_{\mbox{\scriptsize F}}| = |J_{\mbox{\scriptsize A}}| $. The 
quantum spin system in their analysis has the following nearest
neighbor Heisenberg Hamiltonian
\begin{equation}
{\cal H} = \sum_i J_i {\bf S}_i \cdot {\bf S}_{i+1} ,
\end{equation}
with a bond probability distribution which is independent of $i$
\begin{equation}
P(J_i) = p \delta(J_i + J_{\mbox{\scriptsize F}}) 
+ (1-p) \delta(J_i - J_{\mbox{\scriptsize A}}) ,
\end{equation}
where $ 0 \leq p \leq 1 $, and $\delta$ is the Kronecker delta
function. For $ p = 0 $ and 1 we have a purely AF or FM  spin chain,
respectively.

The analysis of this model by
high-temperature expansion \cite{Furusaki} and by a real space
renormalization group (RSRG) scheme \cite{Westerberg} suggests that
three different temperature regimes are present in this system.  In
the high-temperature regime the spins essentially behave as
independent degrees of freedom. If the temperature is lowered down to
$ k_{\mbox{\scriptsize B}} T \sim J $ the spins start to correlate. 
First they align within
the segments of purely FM  or AF bonds. The coupling among the
segments is weak. Thus, the spins in each segment create an effective
single spin degree of freedom which is rather large in the case of FM 
bonds and $ S=0 $ or $ 1/2 $ for AF bonds. At intermediate
temperatures these effective spins behave as independent due to the
thermal fluctuations. However, they begin to correlate at lower
temperatures. The intermediate and low-temperature regimes are
described by a random spin model with a basically continuous random bond
distribution as well as random spin sizes (effective spin sizes).

The limit of $ {\rm T} \to 0 $ was analyzed by Westerberg and
coworkers by means of a RSRG scheme for this type of
model\cite{Westerberg,Westerberg2}. Here we will briefly review their
main results. As the temperature $T$ is lowered, a growing number of
spins is correlated in clusters consisting of $n$ spins on the
average, which form a large spin $S$. The spin size $S$ scales with $S
\propto n^{1/2}$ and the average number of spins in a cluster scales
with the temperature as $n \propto T^{-2 \alpha}$. The scaling
exponent $\alpha \approx 0.22$ was determined numerically 
\cite{Westerberg,Frischmuth}. Because the large effective spins $S$ of
the clusters behave as essentially independent under thermal
fluctuations, one expects a Curie-like susceptibility $ \chi \propto
C/T$ with a $T$ dependent Curie constant $C$.

From this result we conclude that, in principle, this system exhibits
three different temperature regimes, each with its own Curie-like
susceptibility and Curie constant. In the high-temperature regime ($
k_{\mbox{\scriptsize B}} T \gg J $) the Curie-behavior originates from 
the $ S=1/2 $ spins leading 
to $ \chi = \frac{\mu^2_{\mbox{\tiny B}} C}{k_{\mbox{\tiny B}} T} $, where
$\mu_{\mbox{\scriptsize B}}=\frac{e\hbar}{2mc}$ is the Bohr-magneton 
and the Curie constant is $C=S(S+1)/3=1/4$. In the intermediate 
temperature regime ($ k_{\mbox{\tiny B}} T \sim J $) the 
effective spins of the segments give a $ p $-dependent
Curie constant $C\eff$. Finally a crossover to a scaling
regime with a new Curie behavior occurs at very low-temperatures.
Obviously the three Curie constants should be a decreasing sequence as
the number of available degrees of freedom is decreasing with lowering
temperature.

The separation into three regimes is expected to be visible in the
specific heat too. Peak- or shoulder-like structures indicate the
boundaries between the regimes as they are sign of correlation of
degrees of freedom. One boundary occurs 
near $ k_{\mbox{\scriptsize B}} T \sim J$ and the second 
at some lower temperature.  In the scaling regime at
very low temperatures, the assumption of independent large spins $ S $
of correlated clusters leads to $c_{\scr{V}} / T \propto T^{2
  \alpha -1} |\ln T|$ \cite{Westerberg}.

The crossover between the high- and intermediate temperature regime
has been examined by the high-temperature expansion. For low
temperatures, however, only the limiting behavior for $ T \to 0 $ has
been investigated for an effective Hamiltonian with a broad random
distribution of couplings $J_i$ \cite{Westerberg,Frischmuth}, which is in
contrast to the discrete distribution of the initial spin couplings in
\PtIr. From these results we know that the effective spin scaling
regime starts at very low temperatures, which are hard to observe by
experiment. The actual crossover from the intermediate to the
low-temperature regime can be investigated more easily by experiment,
but has not yet been analyzed so far. In this paper we would like to
close this gap by demonstrating that the intermediate temperature
regime is well-defined and observable in both the specific heat and
the susceptibility by an exact treatment of the original Hamiltonian.

The correlations among the FM  bonds in 
\emergencystretch=100pt
\PtIr \mbox{} are another
important aspect which has not yet been examined. We will investigate
the implications of these correlations in a realistic model where the
FM bonds always occur pairwise and study its differences to the
uncorrelated model in the magnetic susceptibility and the specific
heat in an experimentally accessible temperature range. These data
will provide a sensitive test for the randomness of the distribution
of the $\rm Pt$-ions in \PtIr.

Our investigations are based on Quantum Monte Carlo (QMC) simulations
by the loop algorithm \cite{evertz_prl}, which is an excellent method
to simulate accurately the thermodynamics of large spin systems.

%****************************************************************
\section{Numerical Methods}
%****************************************************************
The QMC loop algorithm \cite{evertz_prl} is a finite temperature method
based on the Trotter-Suzuki decomposition \cite{trotter_suzuki} of the
partition function $Z$ and is ideally suited for the calculation of
thermodynamic properties of unfrustrated spin systems. It allows the
direct and exact computation of various thermodynamic observables,
e.g. the uniform magnetic susceptibility $\chi$ or the internal energy
$u$, without introducing any approximations. In contrast to the
classical Metropolis world-line algorithm \cite{metropolis}, the
loop-algorithm does not suffer from prohibitively long
auto-correlation times. Additionally, the introduction of improved
estimators \cite{improved_e} gives a further reduction of the
variance. This enables us to investigate much bigger problems than
with previous QMC methods.

\begin{figure}
\begin{center} \null\
 \psfig{figure=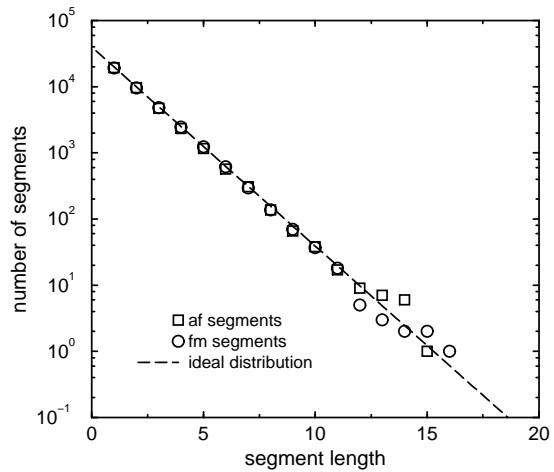,width=\linewidth} \end{center}
 \caption[*]{Statistic of the total distribution of the segments with
 $l_{s}$ FM  (squares) and AF (circles) bonds in the configurations we
 have used for the simulation of the unconstrained model. The dashed
 curve shows the ideal distribution.}  \label{Fig_Bond} 
\end{figure}

For the calculation of the random bond models, we have considered up
to 400 samples of different random bond configurations, each
consisting of a chain of $L=400$ sites with periodic boundary
conditions. In Fig.~\ref{Fig_Bond} we show the distribution of FM  (AF)
segments of length $l_{s}$ in our configuration samples.
We have performed calculations in the temperature range from
$T/J=1/60$ to $T/J=10$. Unless otherwise mentioned, we have used
Trotter numbers between 20 and 120 and extrapolated to Trotter
time-steps $\dt J=0$.  We made $10^{4}$ multi-cluster updates for
thermalization, followed by $2\times 10^5$ up to $2\times 10^6$
multi-cluster updates for measurements, depending on the temperature
$T$ and the Trotter-number.

The value of an observable $\cal O$ is estimated by averaging over $M$
successive measurements $O_{j}$ coming from the QMC
simulation:
$        \langle {\cal O} \rangle \approx
        \overline{O} =
        M^{-1} \sum_{j=1}^{M} O_{j},
$
and the error is determined by the variance of the measurements.
In this way we have measured directly the internal energy $u$, the
total magnetization $M$, and the uniform susceptibility $\chi$ for each
configuration. Due to the efficiency of our QMC algorithm, we have
been able to measure the internal energy $u$ accurately enough to
calculate the specific heat numerically as the first derivative of the
internal energy $u$ with respect to the temperature $T$.
After the calculation of all observables for each individual
configuration, the average over all configurations has been
taken.  In order to get a correct estimate of the errors, we have
taken into account the variance of the observable for the set of
configurations and the error of the observable for a single
configuration. In general, the error-bars for the observables of a
single configuration are much smaller than the variance of the
observables for the set of configurations, and therefore we can
neglect them completely.

%****************************************************************
%%%%%%%%%%%              RESULTS
%****************************************************************
\section{Results}
We have studied two different models. First, we have investigated
a generic random bond model with ${\cal H} = \sum_i J_i {\bf S}_i
\cdot {\bf S}_{i+1}$ and the bond probability distribution
\begin{equation}
P(J_i) = p \delta(J_i + J_{\mbox{\scriptsize F}}) 
+ (1-p) \delta(J_i - J_{\mbox{\scriptsize A}}),
\end{equation}
where $ 0 \leq p \leq 1$ 
and $|J_{\mbox{\scriptsize F}}|$=$|J_{\mbox{\scriptsize A}}|$ for 
the case of $p=0.5$,
i.e., the same probability for finding a FM  as an AF bond. This model
has been studied in a number of papers using different methods
\cite{Furusaki,Westerberg,Westerberg2}. In the rest of this paper 
we will call this model the ``unconstrained'' model.

In the real \PtIr \mbox{} alloys, there are some additional
restrictions. As we have previously mentioned in the introduction, by
replacing the spinless $\rm Pt$-ions with $\rm Ir$-ions, two FM  bonds
are created between the spin $1/2$ carrying $\rm Ir$-ions and its
neighboring $\rm Cu$-ions. In addition to this pairwise correlation of
FM bonds, the FM  couplings are stronger than the AF couplings, an
estimate gives $|J_{\scr{F}}|=4|J_{\scr{A}}|$. By generating 
a random sequence of
$\rm Pt$- an $\rm Ir$-ions we have generated a configurations
equivalent to and \Ir \mbox{} substitution of $x=0.5$, which
corresponds to the probability of finding a FM  bond with $p=2/3$.  We
will refer to these systems as the ``constrained'' model in the
following text of this paper.

%%%%%%%%%%%%%%%%%%%%%%%%%%%%%%%%%%%%%%%%%%%%%%%%%%%%%%%%%%%%%%%%%%%%%%%%
% SUSCEPTIBILITY
%%%%%%%%%%%%%%%%%%%%%%%%%%%%%%%%%%%%%%%%%%%%%%%%%%%%%%%%%%%%%%%%%%%%%%%%
\section{Susceptibility}
Let us consider the unconstrained model first. In the high-temperature
regime $k_{\mbox{\scriptsize B}}T \gg J$, we have independent single 
spins as in the
uniform AF or FM  chain, and a Curie-law for the susceptibility
$\chi_{\scr{cl}}(T)=\mu^{2}/(4 k_{\mbox{\scriptsize B}} T)$ 
for $T \rightarrow \infty$.  As the
temperature $T$ is lowered, the individual spins start to
correlate. In the AF segments, they form the collectively lowest
possible total spin ($S=0$ for an odd number of bonds $l_s$ or $S=1/2$
for an even number $l_s$), while the FM  segments form the collectively
largest spin (by aligning all spins parallel).  Due to the misfit of
the discrete spectra in the AF/FM spin segments (the finite size gap
$\delta E_{\scr{FS}}$ is $\delta E_{\scr{FS}} \propto J/l_s$ for AF
and $\delta E_{\scr{FS}} \propto J/l_{s}^2$ for FM  segments), the
excitations remain localized in the segments and the interactions
among different segments remain very weak. In this intermediate
temperature regime the segments essentially behave as uncoupled
effective spins $S\eff$.  These uncoupled effective spins can be seen
in a second Curie-law of the susceptibility $\chi$ on an intermediate
energy-scale $( \propto
\delta E_{\scr{FS}})$ with an effective Curie-constant $C\eff$
depending on the average size of the effective spins
\[
        \chi = \frac{\mu^{2} C\eff}{k_{\mbox{\tiny B}} T} ,
\]
with $C_{\scr{eff}}=\frac{1}{3} \frac{\langle S_{\mbox{\tiny FM}}^{2}\rangle +
\langle S_{\mbox{\tiny AF}}^{2}\rangle } 
{\langle n_{\mbox{\tiny FM}}\rangle + \langle
n_{\mbox{\tiny AF}}\rangle }$ \cite{Furusaki}, 
where $\langle S_{\mbox{\scriptsize FM/AF}}\rangle $
denotes the effective spin of the FM/AF segment, and $\langle
n_{\mbox{\scriptsize FM/AF}}\rangle $ is the average length of 
a FM/AF segment.  The
spins on the boundary of a segment of pairs of FM spins and a segment of
AF spins always have to be counted
to the AF segment and not to the FM  segment, i.e., the total spin of a
FM segment with a length of $l_s$ bonds is $S_{\scr{tot}}=S(l_{s}-1)$
\cite{Furusaki}.  For the unconstrained model with $p=0.5$ and $S=1/2$ one
obtains $C\eff =1/8$ for a chain of infinite length.
  \begin{figure}
    \begin{center}
      \null\ \hspace{-2mm}
      \psfig{figure=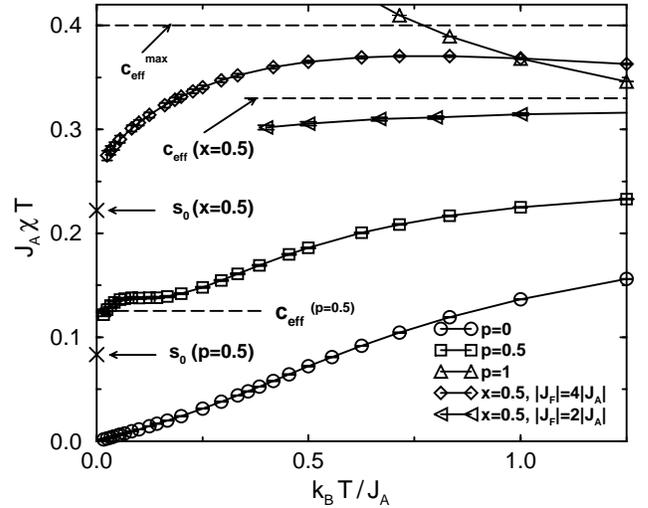,width=1.1\linewidth}
    \end{center}
    \caption[*]{Figure of the
    uniform magnetic susceptibility $\chi$ times temperature $T$. The
    symbols are the results of the QMC simulations for the unconstrained model
    (squares) and the constrained model of \PtIr \mbox{} (diamonds). For
    comparison we also show the results for the uniform FM  and AF
    Heisenberg chain. The Curie-law behavior can be seen by horizontal
    lines, the dashed lines show the results of the statistical
    analysis, $s_0$ are the limits of $\chi T$ for $T\rightarrow 0$.}
    \label{Fig_Susc}
  \end{figure}
 
In Fig.~\ref{Fig_Susc} we show $\chi J T$ as a function of the
temperature, such that the Curie-law behavior of the effective spins
can be seen as a plateau. From our QMC simulations we get $C\eff
=0.138 \pm 0.003$, whereas the value of the effective constant $C\eff$
obtained by statistical analysis of the bond distribution is $C\eff
=0.1252$, and it is shown by a dashed line in the graph.  Previous
results by the transfer matrix method and high-temperature expansions
obtained $C\eff =0.13$\cite{Furusaki}. However, in the
high-temperature expansion, a possible error stems from the
extrapolation of the Pad\'e approximants. There is also some range for 
modification by considering to
which segment a spin on the border between FM  and AF segments has to be
counted in the statistical analysis:
In general, the spins at the border between these segments tend to be
included in the AF segments, but if a long AF segment is next to a
short FM segment, some of these boundary spins rather couple to the FM
segment, resulting in an increase of $C\eff$. A detailed analysis of the
bond distribution used in our QMC calculations shows that if we count
the spins on the border of AF segments consisting of six and more
bonds next to FM segments consisting of three and less bonds to the
FM segment, we rather get $C\eff =0.131$ and similarly for AF
segments consisting of five and more bonds next to FM segments
consisting of three and less bonds we find $C\eff =0.138$ instead, in
agreement with our QMC results.

The temperature range of the regime of uncoupled effective spins is
determined by the extension of this Curie-law plateau, it starts at
$k_{\mbox{\scriptsize B}} T \approx 0.16J$ and 
ends at $k_{\mbox{\scriptsize B}} T\approx 0.06J$. The
region of this regime compares favorably to the simple estimate given
by the finite size gap of the spin segments $\delta E_{\scr{FS}}$.

At very low temperatures, interactions among the effective spins
become relevant and the effective spins start to correlate.  These
interactions are again described by a Heisenberg model with $H\eff
=\sum_i J^i_{\scr{eff}} {\bf S}^i_{\scr{eff}} \cdot {\bf
  S}^{i+1}_{\scr{eff}}$, where the sum $i$ is now taken over the spin
segments. The couplings $J^i_{\scr{eff}}$ are random in both sign and
magnitude, and the size of the effective spins $S_{\scr{eff}}$ is
random as well.  Following ref.~6, the distribution of
the effective spins $S^{i}_{\scr{eff}}$ and their interactions
$J^{i}_{\scr{eff}}$ is very broad. Hence two neighboring spins with
the largest energy gap to their first excited state will be locked
into their ground-state, consisting of their maximal ($J\eff <0$) or
minimal ($J\eff >0$) spin state and form a new effective spin
$\widetilde{S\eff}$. As more and more spins freeze out, the effective
spin size $S\eff$ scales with the number of original spins in the
cluster as $S\eff^{i} \sim n^{1/2}$ (random walk). The average number of
original spins in a cluster $n$ depends on the temperature $T$ and scales
with a power law $n \sim T^{-2\alpha}$, with $\alpha=0.22 \pm 0.01$
\cite{Furusaki}, consistent with the result $\alpha=0.21 \pm 0.01$ of
ref.~8

\begin{figure}
    \begin{center}
      \null\ \hspace{-2mm}
      \psfig{figure=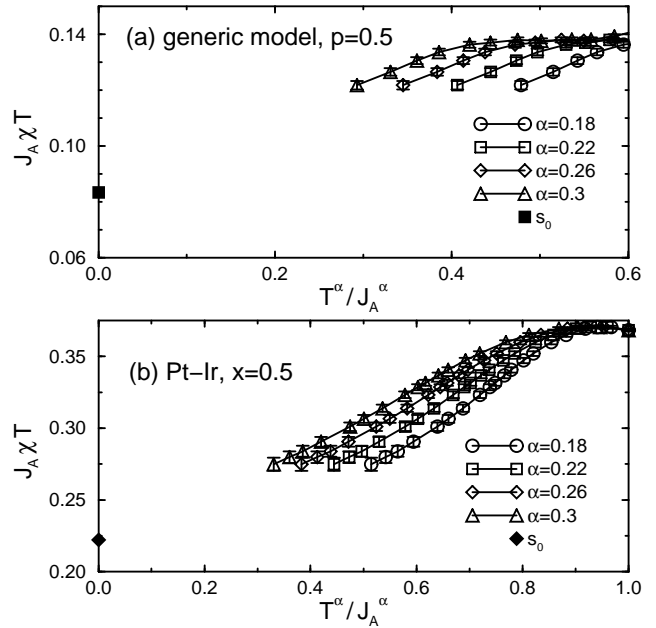,width=1.1\linewidth}
    \end{center}
    \caption[*]{Low
    temperature scaling behavior of $\chi T$ for the un\-constrained model
    (a) and the constrained model (b). The low-temperature scaling
    regime is characterized by $\chi T = s_0 + T^{\alpha} / \gamma +
    O(T^{3\alpha})$. The open symbols denote different values of the
    scaling exponent $\alpha$, the full symbols denote the $T\rightarrow 0$
    limit}
    \label{Fig_Susc_Scaling}
  \end{figure}
In Fig.~\ref{Fig_Susc_Scaling} we have plotted $J \chi
T$ as a function of $T^{\alpha}/J$ for different values of $\alpha$.
The low 
temperature behavior has been calculated in ref.~8:
\begin{equation} \label{eq_lowT_chi}
 \chi T = s_0 + \frac{T^{\alpha}}{\gamma} + O(T^{3\alpha}),
\end{equation}
where $s_{0}$ is the groundstate expectation value of the total spin
per site
\begin{equation} \label{eq_s0}
  s_{0}=\frac{\mu^{2}}{3 k_{\mbox{\tiny B}} L} \left( \sum_{i=1}^{L} \delta_{i}
  S_{i} \right ) ^{2},
\end{equation}
where $\delta_{i+1}=-\delta_{i} \mbox{sgn}(J_{i})$ and
$\delta_1=1$.
The beginning of the low-temperature scaling regime is estimated to start at
$T<0.05 J_0$ for the effective Hamiltonian \cite{Frischmuth}, where $J_0$
is the maximum of a broad random distribution of couplings. This
distribution of couplings corresponds to the effective couplings
$J\eff$ of the intermediate temperature regime. These couplings are
about an order of magnitude smaller than the original couplings
\cite{Furusaki}. In
Fig.~\ref{Fig_Susc_Scaling}, we can see that a linear extrapolation of
the data points with the scaling exponent $\alpha=0.22$ does not yet
yield the correct zero-temperature value $s_0$ and we conclude that the
scaling regime has not been reached for $T=0.02J$. Unfortunately, critical
slowing down in the QMC simulation at low temperatures prevents us from
reaching temperatures as low as $T<0.005J$.

Next we concentrate on the susceptibility of the constrained
model of \PtIr \mbox{}. In this case we have to take into account the
different strength of the FM  and AF interactions. Starting from the
same high temperature limit $\chi/L = \frac{1}{4} J/T$ as in the
generic case, the FM  bonds start to correlate by aligning parallel at
higher temperatures $T$ than the AF bonds, due to the stronger
coupling $|J_{\scr{F}}| = 4 |J_{\scr{A}}|=4J$.  The formation of these
large spins leads to an increase of $J \chi T$ first. Then the AF
spins start to correlate too, but instead of a plateau as in the case
of the unconstrained model we obtain a broad peak 
near $k_{\mbox{\scriptsize B}} T=J$.

If we calculate the average effective spins $\langle S_{\scr{FM/AF}}\rangle
$ as in the previous case, we obtain an estimate of the effective
Curie constant $C\eff =0.33 \pm 0.017$. This value is a lower bound of
the Curie constant, where the spins within the segments are completely
correlated. The effect of the constraint among the FM  bonds in \PtIr
\mbox{} is very important, if we neglect that we would obtain $C\eff
\approx 0.2731$ instead.  An upper bound of $C\eff^{\scr{max}}=0.40
\pm 0.02$ can be obtained by assuming that only the spins in the FM 
segments correlate and the AF segments remain uncoupled due to their
weaker couplings.  From our QMC data in Fig.~\ref{Fig_Susc_Scaling} we
get $C\eff \approx 0.36$.  The deviation from $C\eff$ of the spin
segment statistic is larger than for the unconstrained model, since
the difference in the coupling strength favors quantum effects on the
boundaries of large FM and small AF segments and makes the simple
statistical estimate more subtle. The effects causing an increase of
of $C\eff$ are twofold in this case: the first is the same as for the
unconstrained model, spins on the border of large AF segments next to
a short FM segments tend to couple to the FM segment. A detailed
analysis of the spin distribution gives $C\eff=0.34$ if we count the
spins on the border of AF segments consisting of five and more bonds
next to FM segments consisting of four and less bonds to the FM
segment, and $C\eff=0.345$ for AF segments consisting of four and more
bonds next to a segments consisting of three and less bonds. The
second reason for a larger value of $C\eff$ is that due to the
relatively weaker AF couplings than the FM couplings, some of the AF
segments are not yet completely correlated. Thus especially the long
AF segments still consist of some original spin-$1/2$ degrees of
freedom, giving rise to a further increase of $C\eff$ as we have
demonstrated for the case of completely uncorrelated AF segments
before.

Additionally we have also determined $C\eff$ by a QMC simulation for
the same distribution as above but with different coupling strength
$|J_{\scr{F}}|=2|J_{\scr{A}}|=2J$. Here we can see again how 
the susceptibility $\chi T J$ increases from the high-temperature 
limit $\chi T J = 1/4$; but in
contrast to the previous case it never reaches the lower bound of
$C\eff =0.33 \pm 0.017$ of the regime of uncoupled effective spins.
In this case, some spin segments already start to correlate and form a
collective spin before the regime of independent effective spins is
established.

In order to investigate the low-temperature scaling behavior of
eq.~(\ref{eq_lowT_chi}) for the constrained model, we have to calculate
$s_0$ from eq.~(\ref{eq_s0}). Let us consider a chain of $L$ $\rm{Cu}$-
ions. There are $Lx$ spin $1/2$ carrying $\rm{Ir}$-ions in the chain
and the ground state expectation value of the total spin is: $\langle
( \sum_{\scr{spins}} S_i )^2 \rangle = (1+x) L \langle S_i^2\rangle +
\sum_{\scr{spins}} \langle S_i S_j \rangle $, where the sum is taken
over all \Cu- and \Ir-ions. For the evaluation of $\sum_{\scr{spins}}
\langle S_i S_j \rangle $ we divide the sum into four terms using the
following notation: we enumerate all \Cu-ions with the index $i$,
while we use the index $\hat{i}$ for the \Ir-ion between the \Cu-ions
$i$ and $i+1$ and write \emergencystretch=100pt 
$\sum_{\scr{spins}}\langle S_i S_j \rangle = 2
\left( \sum_{i<j} \langle S_i S_j \rangle + \sum_{i<
    \hat{j}}\langle S_{i} S_{\hat{j}} \rangle +\sum_{\hat{i}<j} \langle
  S_{\hat{i}} S_{j} \rangle \right. \\ \left. +\sum_{\hat{i}<\hat{j}} \langle
  S_{\hat{i}} S_{\hat{j}} \rangle \right) $. Now we can calculate the
expectation value $\langle S_i S_j \rangle $ by taking into account
that an \Ir-ion exists only with the probability $x$ and that each
missing \Ir-ion changes the sign of the next spin:
\begin{subeqnarray}
  \langle  S_i S_{i+n} \rangle \! \!\!\! & = &  \!\! \!\!
  \frac{1}{4} \sum_{m=0}^n x^m (1-x)^{n-m} (-1)^{n-m} 
  \left( \!\! \! \begin{array}{c} n \\ m \end{array} \!\! \! \right) 
\nonumber \\ \!\!\!\! &  =&\! \! \!\!
   \frac{1}{4} (2x -1)^n \\
  \langle  S_i S_{\widehat{i+n}} \rangle \! \! \!\!  & = &  \! \! \!\!
  \frac{x}{4} \sum_{m=0}^n x^m (1-x)^{n-m} (-1)^{n-m} 
  \left( \!\! \! \begin{array}{c} n \\ m \end{array}  \!\!\! \right)  
\nonumber \\ \!\!\!\! & = & \!\!\!\!
   \frac{x}{4} (2x -1)^n \\
  \langle  S_{\hat{i}} S_{i+n} \rangle  \! \! \!\! & = & \! \! \!\!
  \frac{1}{4} \sum_{m=0}^{n-1} x^m (1-x)^{n-1-m}
(-1)^{n-1-m} 
  \left( \!\! \! \begin{array}{c} n-1 \\ m \end{array}  \!\! \! \right) 
\nonumber \\ \!\!\!\! &  = & \!\!\!\!
   \frac{1}{4} (2x -1)^{n-1} \\
  \langle  S_{\hat{i}} S_{\widehat{i+n}} \rangle \! \!\!\! & = &  \!\! \!\!
  \frac{x}{4} \sum_{m=0}^{n-1} x^m (1-x)^{n-1-m}
(-1)^{n-1-m} 
  \left( \!\!\! \begin{array}{c} n \\ m \end{array}  \!\!\! \right) \nonumber 
\\ \!\!\!\! &  = & \!\!\!\!
   \frac{x}{4} (2x -1)^{n-1},
\end{subeqnarray}
where $\left( \! \begin{array}{c} n \\ m \end{array} \! \right)=n!/(m!(n-m)!)$.
From this we obtain
\begin{eqnarray}
\left\langle \left( \sum_{\scr{spins}} S_i \right)^2 \right\rangle  &= &
 L(1+p)\langle S_{i}^2 \rangle + \nonumber \\
&&\hspace*{-2cm}  \frac{2L}{4}\left[
\sum_{n=1}^{L}(2x -1)^n + \sum_{n=0}^{L}x(2x -1)^n \right. \nonumber \\
&&\hspace*{-1cm}\left. + x \left(\sum_{n=1}^{L}(2x -1)^{n-1} 
+ \sum_{n=1}^{L}x(2x
-1)^{n-1}\right) \right]  \nonumber \\
&=& L,
\end{eqnarray}
which gives in the groundstate $s_0=\langle (
\sum_{\scr{spins}} S_i )^2 \rangle/(3 k_{\mbox{\scriptsize B}} (1+x) L)
=2/9 \approx 0.222222\ldots$.

In Fig.~\ref{Fig_Susc_Scaling} we can see that also for this model the
value we get for $s_0$ is too small if we extrapolate our data
linearly with the correct scaling exponent $\alpha \approx 0.22$. Thus
we conclude that the low-temperature scaling regime starts 
below $T/J_{\mbox{\scriptsize A}}<1/100$.

%%%%%%%%%%%%%%%%%%%%%%%%%%%%%%%%%%%%%%%%%%%%%%%%%%%%%%%%%%%%%%%%%%%%%%%%
% SPECIFIC HEAT
%%%%%%%%%%%%%%%%%%%%%%%%%%%%%%%%%%%%%%%%%%%%%%%%%%%%%%%%%%%%%%%%%%%%%%%%
\section{Specific Heat}
\begin{figure} 
\begin{center} \null\ \hspace{-2mm}
  \psfig{figure=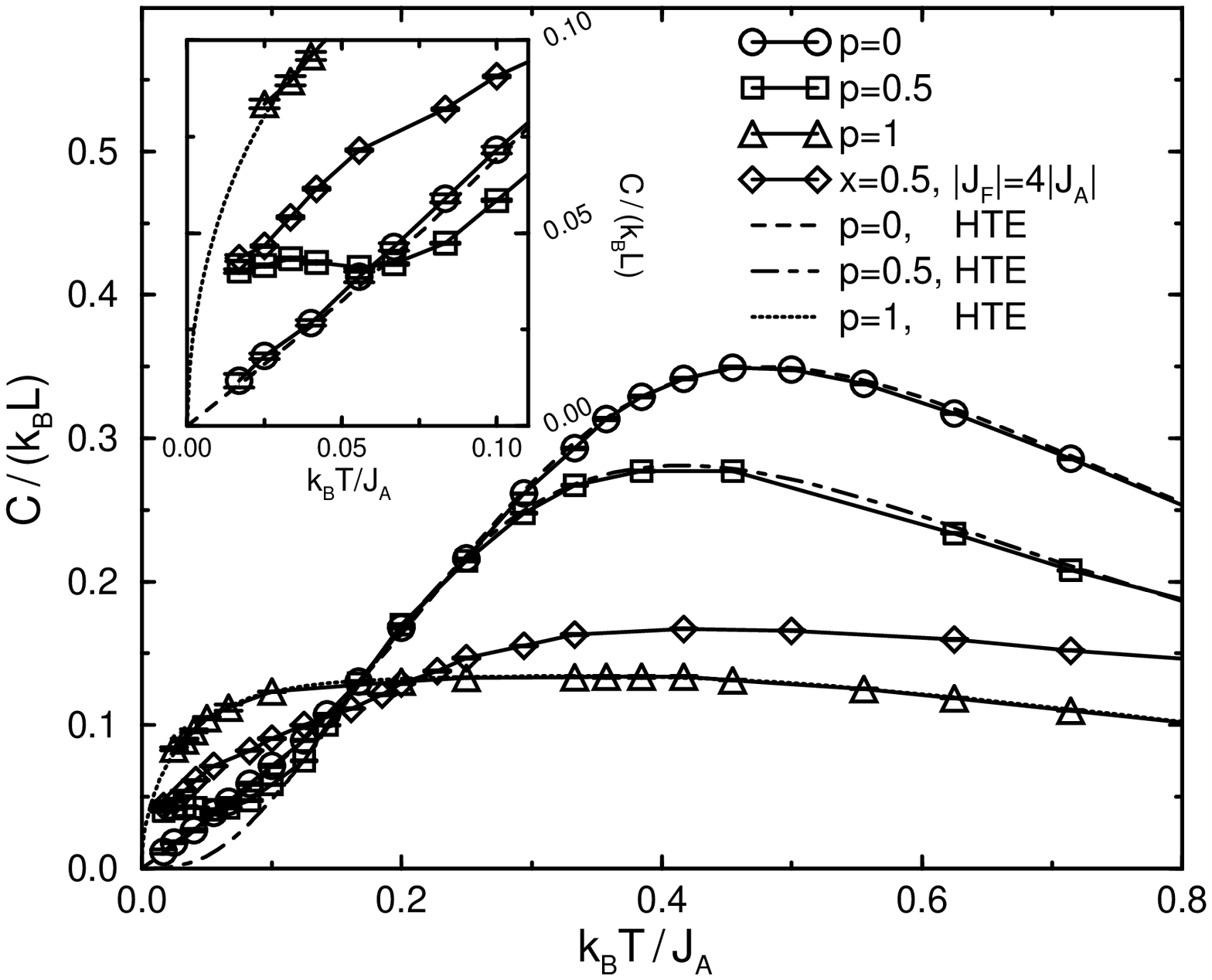,width=1.1\linewidth}
  \end{center} 
\caption[*]{ The specific heat per spin $c_{\scr{V}}$ of the
  unconstrained random bond model and the \PtIr-model.  For
  comparison, we have also shown the results of the uniform FM  and AF
  Heisenberg chain.  The symbols are the results of the QMC simulation
  for the generic model (squares) and the constrained model of \PtIr \mbox{}
  (diamonds), dashed and dotted lines (HTE) are the
  high-temperature expansion results obtained by the extrapolation of
  the two-point Pad\'e approximants \cite{Furusaki}. In the inset,
  one can see the peak in the specific heat of the unconstrained model and a
  cusp for the constrained model which indicate the onset of
  correlations among the effective spins.}  \label{Fig_CV}
  \end{figure}

Again, let us concentrate on the unconstrained model first. In the specific
heat data, the crossover between different temperature regimes should
be visible by peak-like structures, where the spins start to
correlate. In the uniform FM  or AF Heisenberg systems, one broad peak
below $k_{\mbox{\scriptsize B}} T /J \sim 1$ appears, where the 
individual spins start to
correlate. Two peaks are expected in the random system. One where the
original $S=1/2$ spins start to correlate (as for the uniform systems),
and a second one where the segments of the effective spins start to
correlate (at the end of the 
Curie-plateau $k_{\mbox{\scriptsize B}} T/J \sim 1/20$). In
Fig.~\ref{Fig_CV} we show the specific heat per spin $c_{\scr{V}}$ of the
random bond systems and of the uniform AF and FM  chain for comparison.
The large and broad peak is the signature of the correlation of the
individual spins. In the inset, one can clearly see the second peak at
low temperatures of the correlation of the effective spins for the
unconstrained model. This confirms the clear separation of the
different regimes and could not be investigated with the previous
methods, e.g. the high-temperature expansion.

\begin{figure}
    \begin{center} \null\ \hspace{-2mm}
    \psfig{figure=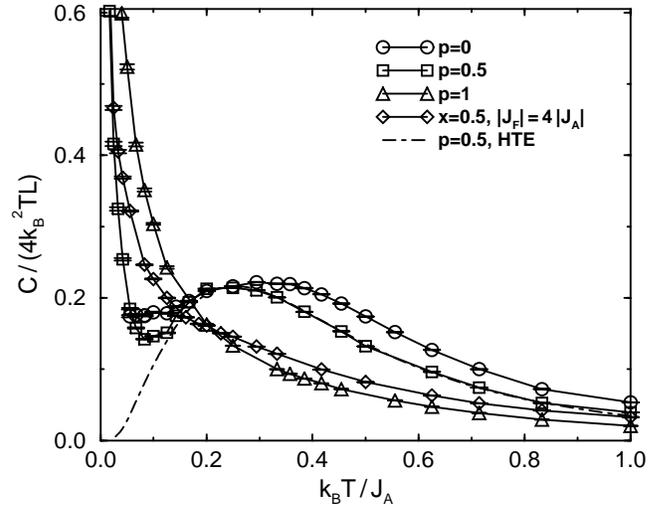,width=1.1\linewidth} \end{center}
    \caption[*]{The specific heat per spin $c_{\scr{V}}$ divided by the
    temperature $T$ for the uniform and random bond systems.  The area
    below the curves corresponds to the entropy $S$. The symbols are
    the QMC results, the dot-dashed line (HTE) is the high-temperature
    expansion result obtained by the extrapolation of the two-point
    Pad\'e approximants \cite{Furusaki}. One can see that the QMC
    and high-temperature expansion results are in perfect agreement
    down to $k_{\mbox{\scriptsize B}} T \approx 0.2 J$, then 
    the QMC results show the
    correct crossover to a divergent behavior.}  \label{Fig_CV_dT}
    \end{figure}

In the scaling regime at very low temperatures $T$, the assumption of
independent large spins $S$ of correlated clusters \cite{Furusaki}
leads to the entropy per site $\sigma \propto \ln ( 2 S(T) +1) / n(T)
\propto T^{-2 \alpha} | \ln T |$. From this one finds $c_{\scr{V}}/T \propto
T^{-2 \alpha -1} | \ln T |$.  But as for the magnetic susceptibility,
we can not reach such low temperatures as to determine the scaling
exponent $\alpha$ from our data.  By the relation $\int_0^{T} \mbox{d}T
c_{\scr{V}}/T$, we can see from the divergence of $c_{\scr{V}}/T$ as $T
\rightarrow 0$ that a large fraction of the entropy is at very low
temperatures, due to the broad spectrum of energy-scales of the
effective spins.  In Fig.~\ref{Fig_CV_dT} we show $c_{\scr{V}}/T$, the area
below the curve corresponds to the entropy $S$. If we consider the
data in Fig.~\ref{Fig_CV_dT} in more detail, we can see that the value
of $c_{\scr{V}}/T$ for the generic model is always smaller 
than $c_{\scr{V}}/T$ 
of the uniform AF Heisenberg chain. Only at very low temperatures, we
can see a sharp crossover to the divergent low-temperature
behavior. This crossover is due to the very different energy-scales of
the effective spins and can not be seen by the high-temperature
expansion.  

More important for comparisons with experiments is the specific heat
of the constrained model of \PtIr. Here, the spins start to
correlate at different temperatures $T/J$ due to the different
coupling strengths. Hence there is no clear transition and the peak
at $T/J \approx 1$ is much weaker than for the unconstrained model.  Also
the peak at the transition to the scaling regime is much weaker, but
we can interpreted the cusp-like structure near $T \approx 0.05J$ as
the corresponding peak in the specific heat. In Fig.~\ref{Fig_CV_dT},
where we show $c_{\scr{V}} /T$, one finds that there is 
no peak in $c_{\scr{V}}/T$
before the crossover to the divergent low-temperature behavior as for
the unconstrained model, but $c_{\scr{V}}/T$ is rather 
continuously increasing for $T \rightarrow 0$.

%%%%%%%%%%%%%%%%%%%%%%%%%%%%%%%%%%%%%%%%%%%%%%%%%%%%%%%%%%%%%%%%%%%%%%%%
% SPECIFIC HEAT
%%%%%%%%%%%%%%%%%%%%%%%%%%%%%%%%%%%%%%%%%%%%%%%%%%%%%%%%%%%%%%%%%%%%%%%%
\section{Conclusions}
We have investigated numerically the thermodynamic properties of one
dimensional spin chains with random FM  or AF couplings. We have
considered an unconstrained model with equal magnitude of the AF and
FM couplings and a realistic model of \PtIr \mbox{} (constrained
model). The thermodynamic properties of the random bond systems are
determined by three different energy scales: at high temperatures, the
energy scale is given by the original spin-spin couplings.  The
susceptibility obeys a Curie law of free spins at very high
temperatures ($T \gg k_{\mbox{\scriptsize B}} T$). By lowering 
the temperature, the spins
start to correlate within the AF and FM  segments and form effective
spins. At intermediate temperatures, these effective spins interact
very weakly, essentially as free spins. The crossover of the
individual spins to effective spin segments can be seen by a peak in
the specific heat and a Curie-law of the magnetic susceptibility of
the effective spins.  The energy scale of this intermediate
temperature regime is given by the interactions of these effective
spins. At still lower temperatures, the interactions among the
effective spins become relevant and the spin segments gradually freeze
out into clusters of correlated spin segments. Also this crossover to
the low-temperature scaling regime is visible by a peak in the
specific heat, although this peak is much smaller than for the
previous transition.

Our results represent the first exact treatment of the original Hamiltonian
showing a clear separation into three different temperature regimes
for the unconstrained model. This separation can be seen by two marked
peaks in the specific heat and two Curie laws in the uniform magnetic
susceptibility at high and intermediate temperatures. In order to
allow the comparison with experimental results, we investigate for the
first time a realistic model of \PtIr \mbox{} with a constrained
distribution of the FM  bonds exactly.

We find that the three energy
levels are not so clearly separated in this case. Because of the
different magnitude of the FM  and AF couplings, the energy scales are
overlapping and the crossover between the three regimes is continuous.
Therefore the peak originating from the correlation of the original
spins is very broad and the onset of correlations among the effective
spins can only be seen by a small cusp in the specific heat. Instead
of a clear plateau in $\chi T$, the effective spins form a broad peak.

Overlapping energy-scales can be found in many random bond systems,
indeed, the formation of a clearly distinguishable intermediate
temperature regime depends subtly on the distribution and strength of
the FM/AF bonds.  However, the general picture still applies also for
systems with overlapping energy scales and we have given upper and
lower bounds for the Curie-constant of the susceptibility in the
intermediate temperature regime of the constrained model.  The
calculation of these Curie-constants can be used as a universal tool
for the analysis of the bond-distribution, e.g. much larger
Curie-constants will result in experiments if the distribution of
\Pt-\mbox{} or \Ir-ions is not really random because of large FM  
clusters which can enhance the Curie-constant drastically at low
temperature. The calculation of 
these Curie-constants can also be applied in the statistical
cluster-analysis of bond-distributions \cite{Frischmuth,Frischmu2}.

We have given upper bounds for the beginning of the low-temperature
scaling regime, $T=0.02J$ for the generic model and $T=0.01J_{\scr{A}}$ for
the realistic model. But despite the very efficient QMC algorithm, we
have not been able to reach the low-temperature scaling regime with
the original Hamiltonian for both models. For the original
Hamiltonian, the scaling regime starts at extremely low temperatures,
such that a completely different approach will be needed to reach this
regime with the original Hamiltonian. However, our numerical results
confirm that the intermediate temperature regime is well described by
an effective Hamiltonian for the effective spins, providing a firm
basis for the investigations of the scaling regime based on the effective
Hamiltonian.

We would like to finish by mentioning that an analogue of the random
bond spin chains exists for very dilutely randomly depleted Heisenberg
ladders \cite{depleted_ladders}. Due to the spin gap in the parent
material, the Curie law of the effective spins can be observed
experimentally more easily and the value of the Curie constant is in
agreement with the theoretical value \cite{exp_ladder}.

%****************************************************************
\acknowledgements
%****************************************************************

We wish to thank B. Frischmuth, A. Furusaki, P.A. Lee, N. Nagaosa,
K. Tanaka, and M. Troyer for helpful discussions.  B.A. gratefully
acknowledges support from an ETH-internal grant No.
9452/41-2511.5. All the calculations have been performed on the Intel
Paragon XP/S-22 MP of ETH Z\"{u}rich.

%****************************************************************

\end{document}